\documentclass[twocolumn,amsmath,amssymb,aps,prd,longbibliography]{revtex4-2}
\usepackage{graphicx}
\usepackage{upgreek}
\usepackage{color}
\usepackage{natbib}
\usepackage{xcolor}
\usepackage{array}
\usepackage{multirow}
\usepackage{comment}
\usepackage[colorlinks=false, linkbordercolor=red, citebordercolor=green, urlbordercolor=cyan, pdfborderstyle={/S/U/W 1}]{hyperref}

\usepackage{soul}
\usepackage{cancel}
\begin{document}

\title{Emergence of super-Poissonian light from indistinguishable single-photon emitters}
\author{A.~Kovalenko$^{1}$,  D.~Babjak$^{1}$, A.~Le\v{s}und\'{a}k$^{2}$, L.~Podhora$^{1}$, L.~Lachman$^{1}$, P.~Ob\v{s}il$^{1}$, T.~Pham$^{2}$, O.~\v{C}\'{i}p$^{2}$, R.~Filip$^1$, and L.~Slodi\v{c}ka$^{1}$}
%\email{slodicka@optics.upol.cz}

\affiliation{$^1$ Department of Optics, Palack\'{y} University, 17. listopadu 12, 771 46 Olomouc, Czech Republic \\
$^2$ Institute of Scientific Instruments of the Czech Academy of Sciences, Kr\'{a}lovopolsk\'{a} 147, 612 64 Brno, Czech Republic\\}

\date{\today}

\begin{abstract}
The optical interference constitutes a paramount resource in modern physics. At the scale of individual atoms and photons, it is a diverse concept that causes different coherent phenomena. We present the experimental characterization of both coherent and statistical properties of light emitted from ensembles of trapped ions increasing with a number of contributing phase-incoherent independent atomic particles ranging from a single to up to several hundreds. It conclusively demonstrates how super-Poissonian quantum statistics non-trivially arises purely from the finite number of indistinguishable single-photon emitters in the limit of a single detection mode. The achieved new optical emission regime in which these independent atoms contribute coherently to the super-Poissonian statistics provides a new perspective on the emergence of optical coherence at the atomic scale and constitutes a unique toolbox for its generation and control at the most microscopic level.
\end{abstract}

\maketitle

%%%%%%%%%%%%%%%%%%%%%%%%%%%%%%%%%%%%%%%%%%%%%%%%%%%%%% Introduction
%\section{Introduction}

The generation of classically fully coherent light from macroscopic lasers has become a rudimentary task with many applications indispensable in our everyday life. On the other hand, at the level of individual atoms or quantum emitters, optical interference of emitted individual photons is very diverse and supports many unexplored quantum phenomena. At this microscopic scale the optical coherence has been dominantly explored in two extreme limits. Experiments with large atomic clouds provided fundamental insight into the macroscopic limit of optical coherence on collective interactions with light, implementation of nonlinear couplings, or on possibilities of control of atomic motion~\cite{hammerer2010quantum,bromley2016collective,jennewein2016coherent,vsantic2018nonequilibrium,guerin2017light,ortiz2019mollow}. However, the limit of the large thermal atomic systems could not simultaneously provide tools for a direct observation of the source of optical coherence or its emergence from the contributions of individual atoms. In the opposite regime of interaction of light with a small and well defined number of emitters, the crucial role of optical coherence was dominantly manifested in different fundamental properties of atomic emission and its statistical behaviour with single trapped atoms or ions~\cite{diedrich1987nonclassical,eschner2001light,dubin2007photon,leuchs2013light,slodivcka2013atom,wolf2016visibility,gerber2009intensity,ourjoumtsev2011observation}. The scalability of the underlaying coherent aspects to several individual atomic scatterers~\cite{diedrich1987nonclassical,eichmann1993young,gomer1998single,wolf2016visibility,obvsil2019multipath,reimann2015cavity,wolf2020light} already stimulated applications including optical generation of entanglement of atoms~\cite{slodivcka2013atom,araneda2018interference} or spatial imaging~\cite{richter2021imaging}. Despite microscopic descriptions of the crucial coherent phenomena in a large atom number limit almost exclusively assume the coherent contributions from individual emitters~\cite{hammerer2010quantum,bromley2016collective,jennewein2016coherent,vsantic2018nonequilibrium,guerin2017light,ortiz2019mollow}, the corresponding underlaying emergence of coherence even in the most elementary regime of statistically independent individual single-photon emitters remains fully unexplored. It manifests in generation of super-Poissonian light only from indistinguishable emissions of the stable number of single-photon emitters, expected since the beginning of quantum optics~\cite{mandel1965coherence,paul1982photon,loudon2000quantum}. This is mostly due to challenging requirements on the efficient optical detection of light emitters which are mutually distant on the scale of optical wavelengths and the simultaneous feasibility of their fundamentally indistinguishable detection corresponding to a single spatial optical mode with sufficient photon rate.

The first-order coherence of light upon scattering from finite ensembles of individual atoms can be directly accessed by observation of the interference patterns in an analogy to a Young's slit experiment with several trapped and laser cooled ions~\cite{eichmann1993young,araneda2018interference,wolf2016visibility,obvsil2019multipath}. Their visibility is mostly limited by the residual portion of inelastically scattered light and thermal atomic motion. It has been concluded that, while there exist in principle scalable configurations with respect to the number of participating atoms $N$~\cite{obvsil2019multipath}, it requires a detailed knowledge of many system parameters to access the actual degrees of the first-order coherence relevant for applications, which can otherwise remain uncertain in the case of large $N$. %\textbf{In addition, the bare observability of the first and second order coherence properties of light scattered from many atoms has been so far studied separately, solely in the mutually exclusive experimental regimes.}
On the other hand, it could not be directly observed in realizations capable of detection of intensity correlations from a large and stable number of independent atoms as these were so far measured solely in the spatially multi-mode detection configurations~\cite{obvsil2018nonclassical,richter2021imaging}, which provided a crucial enhancement of the probability of detection of the multi-photon events in the limit of the realistic overall detection efficiencies. The corresponding initial implementations in the single-mode detection limit employed coupling of the atomic fluorescence to a high finesse cavity or optical nanofibers~\cite{hennrich2005transition,nayak2009antibunching}, as it was predicted that the fundamental free-space demonstration is too challenging. However, the crucial condition on the independence of emitters must be carefully considered in these, in principle efficient, coupling scenarios. Moreover, these experiments included fluctuating - Poissonian atom number distributions affecting the interference. %However, it is the bare feasibility of a single-mode detection of intensity correlations which constitutes a crucial and indispensable condition for controllable microscopic tests of a large number of collective scattering phenomena in atomic physics and in their direct applications in optical sensing or long-distance quantum communication~\cite{bromley2016collective,jennewein2016coherent,ficek2002entangled,pleinert2017hyperradiance,thiel2007quantum}.

Here, we present the experimental characterization of coherent and statistics features of light scattered from stable ensembles of noninteracting independent single-photon emitters represented by trapped ion crystals formed in a linear Paul trap. We study the second-order coherence of the light scattered in a free space and its dependence on the number of independent emitters. We reach the unprecedented experimental regime where the detected photon indistinguishability is guaranteed by a single-mode detection and, simultaneously, the detected photon rate is sufficient for the observation of two-photon contributions necessary for the evaluation of the intensity correlation. We observe unambiguous manifestations of rising super-Poissonian nature of the scattered light from up to several hundreds of independent atomic emitters in the measurement and analysis of a photon statistics in a single-mode photon counting experiment. As the normalized correlation functions are loss independent, we simultaneously analyse statistical features of the interfering light using complementary approach~\cite{lachman2016nonclassical,moreva2017direct,obvsil2018nonclassical,qi2018multiphoton}, independent of Poissonian background noise.  Both approaches complementary reveal that the single mode super-Poissonian statistical behaviour could not be attributed to the increased losses or addition of signal with Poissonian statistics.

%\section{Indistinguishable emission from independent atomic scatterers}

\begin{figure*}[th!]
\begin{center}
\includegraphics[width=1.6\columnwidth]{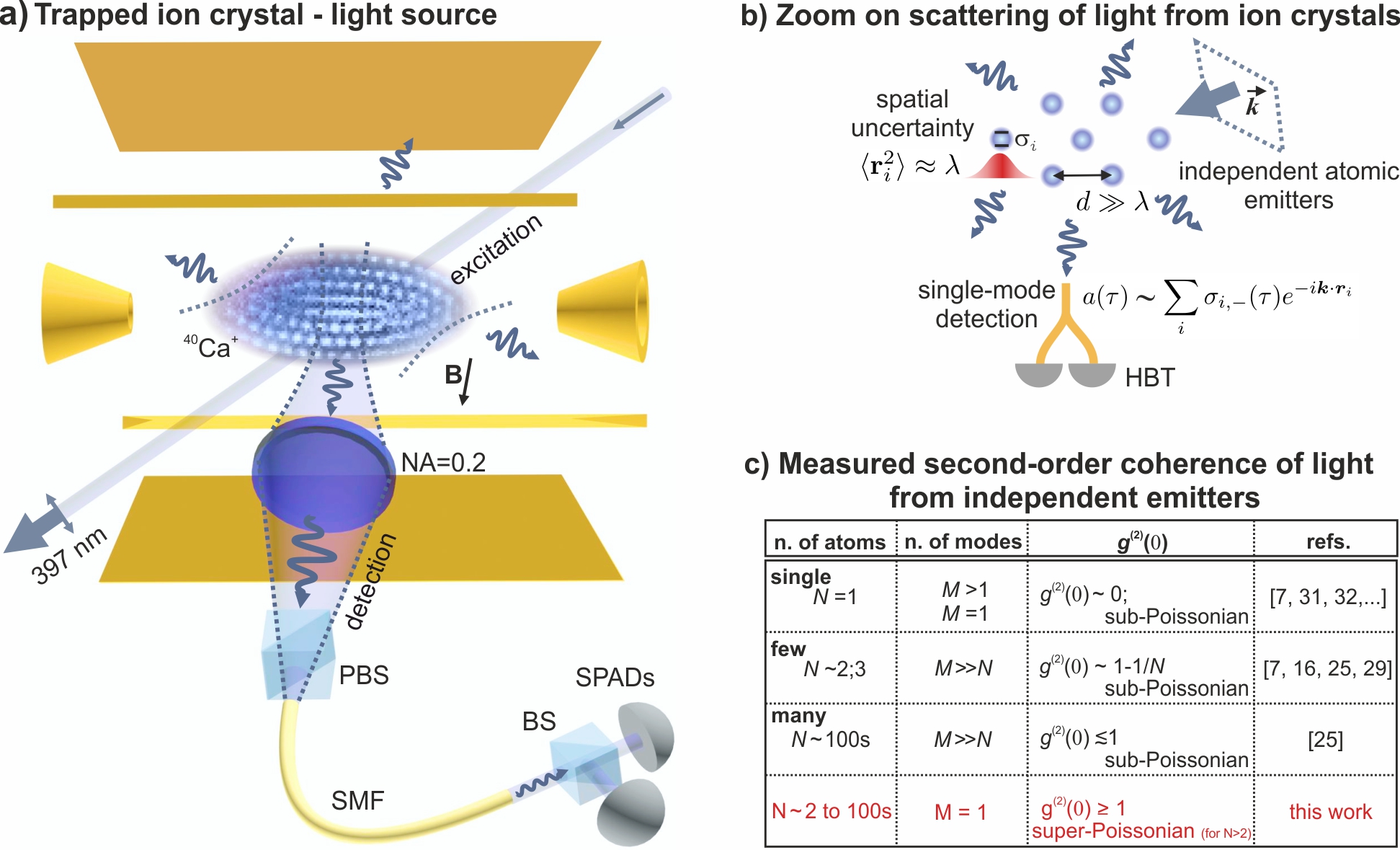}
\caption{a) A simplified scheme of the employed experimental setup. A large but finite and temporarily stable ensembles of non-interacting, statistically independent, and phase-randomized atoms are implemented as crystallized structures of $^{40}{\rm Ca}^{+}$ ions confined in the linear Paul trap. The light scattered from the excitation 397~nm laser beam is collected using an optical objective with numerical aperture of~0.2 and coupled into a single-mode optical fiber. A weak magnetic field is set along the observation direction to lift the degeneracy of the energy manifolds participating in the employed 4$^2{\rm S}_{1/2}\leftrightarrow 4^2{\rm P}_{1/2}\leftrightarrow 3^2{\rm D}_{3/2}$ energy level scheme. The polarization filter consisting of a quarter waveplate (QWP) and a polarization beam splitter (PBS) is set to transmit only one $\sigma$ polarization. The intersection of the detection optical mode with the excitation laser determines the effective number of ions contributing to the detected fluorescence signal. The second-order optical coherence quantified by $g^{(2)}(\tau)$ is accessed using the measurement of photon statistical properties on the set of two single-photon avalanche diodes (SPADs) arranged in a Hanbury Brown and Twiss (HBT) configuration with a balanced beam splitter (BS). The schematic drawing in b) illustrates the scattering scenario, where the input laser beam with a wave vector $\vec{k}$ scatters off atoms, with part of the scattered light reaching the single-mode detection setup.  The table in~c) summarizes the available regimes of measurement of second-order coherence at zero time delay with individual independent emitters. The well understood and experimentally thoroughly explored case of a single emitter $(N=1)$ provided numerous confirmations of anti-bunched and strongly sub-Poissonian character of emitted light with recent values of $g^{(2)}(0)\sim 10^{-4}$~\cite{higginbottom2016pure,crocker2019high}. This limit does not depend on the number of detection modes $M$ and employed spatial filters served mostly for the suppression of background light noise. The possibility of the controllable addition of a small number of emitters in the regime of in principle distinguishable - spatially multi-mode detection resulted in the sub-Poissonian signal in an agreement with the expected multi-mode scaling of $g^{(2)}(0)$ with $N$~\cite{loudon2000quantum,diedrich1987nonclassical,gomer1998single,obvsil2018nonclassical,moreva2017direct}. The regime of a large $N$ provided recently first confirmation of the fundamental feasibility of preservation of the nonclassical sub-Poissonian character of light from a stable finite number of single-photon emitters~\cite{obvsil2018nonclassical}. The corresponding single-mode detection limit remained unexplored despite its paramount importance for the collective coherent phenomena and its fundamental role in understanding the emergence of second order coherence at the level of individual independent emitters.  %when comparing the multi-mode detection of fluorescence from a single ion ($N_{\rm tot}=1$, $M \gg 1$), from many ions ($N_{\rm tot}=275\pm 1$, $M \gg 1$), and from similarly large ion trapping crystal in the single-mode detection setup ($N_{\rm tot}=288 \pm 50 $, $M = 1$). The coincidence time window has been set to 1~ns. We note that the different amount of noise in the presented three measurements reflects the corresponding large difference in the detected photon rates when comparing single to multi-mode observation regimes and single to many ion light sources.
}
\label{fig:block_scheme}
\end{center}
\end{figure*}

The first-order interferences correspond to the fundamental impossibility of assigning the emission from a particular atom to the detected photon~\cite{mandel1991coherence,ficek2005quantum,araneda2018interference}. When scaling up the number of independent atomic scatterers, the photon path indistinguishability of the scattered light is typically achieved by employment of the single spatial mode detection setups. However, the requirement of \emph{non-interacting atoms} necessitates large interatomic distances $d \gg \lambda$ and small optical depths of the atomic sample in order to effectively suppress dipole-dipole couplings and observability of the corresponding collective scattering phenomena~\cite{araujo2016superradiance}. While this is naturally satisfied for crystals formed by laser cooled ions in Paul traps~\cite{devoe1996observation,mortensen2006observation}, simultaneous achievement of the sufficient photon rate from many independent ions is compromised by the effective solid angle fraction of the collection optics which can efficiently couple light from different atoms into the same single spatial mode. In conventional experiments with ions of alkaline earth metals including the $^{40}$Ca$^+$ ion employed here, efficient collection is achieved using objectives with high numerical apertures, which results in count rates of light emitted on optical dipole transitions that are sufficient for measuring the photon correlations. However, projection on a single spatial mode from realistic radially aligned linear or 2-d ion crystals with such numerical apertures on the order of ${\rm NA} \approx 0.1$ already results in a spatial resolution at the ion position on the order of $1/5\,\lambda$ which corresponds to the achievable inter-ion distances or smaller~\cite{devoe1996observation}. While one natural feasible solution corresponds to the observation of atoms aligned along the detection optical axis~\cite{obvsil2019multipath}, the employed linear trap in this reference offers insufficient optical access corresponding to the numerical aperture ${\rm NA} \ll 0.1$ in the axial direction for the efficient measurement of photon correlations in the intermediate ion number limit $N \sim 2 - 10$. We thus realize a specific arrangement with large 3-d ion crystals, which allows for the detection of up to several tens of ions along the radial observation direction and their simultaneous coupling into a single spatial mode. The schematic drawing of the employed experimental setup is shown in the Fig.~\ref{fig:block_scheme}-a). The details of the experimental setup can be found in the reference~\cite{obvsil2018nonclassical} and here we summarize only main features and relevant differences to this and other setups~\cite{richter2022collective} necessary for understanding the presented measurements in a single-mode regime. A linear Paul trap is employed for the localization of $^{40}$Ca${^+}$ crystals with total numbers of ions ranging from $N_{\rm tot}=1$ to up to several hundred. Ions are confined by a combination of DC potentials $U_{\rm tip}= 800 {\rm\ V}$ applied to the two hollow axial electrodes and radio-frequency electric field applied to radial electrodes at $\omega_{\rm RF}=(2 \pi)\times 30$~MHz. The resulting radial and axial secular frequencies for a single trapped ion are $\nu_x \approx \nu_y \approx (2\pi)\times 2$~MHz and $\nu_z \approx (2\pi)\times 1.2$~MHz, respectively. The ions are Doppler cooled using a 397~nm laser beam red-detuned from the $4^2{\rm S}_{1/2}\leftrightarrow 4^2{\rm P}_{1/2}$ transition propagating in the plane perpendicular to the observation direction. We note that in the presented experiments, the 397~nm laser detuning has been particularly set such that the residual thermal motion of ions is sufficient for \emph{randomization of the relative phase of the light} scattered from different atoms.%, which corresponded to about $\Delta_{397}\approx$ \textbf{-XX}~MHz from the resonance.
The population of the metastable $3^2$D$_{3/2}$ manifold is reshuffled back to the cooling transition using 866~nm laser light. The degeneracy of Zeeman states is lifted by the application of a static magnetic field with a magnitude $|\vec{B}|=10$~Gauss along the observation direction. Both excitation laser polarizations are set to linear with a direction perpendicular to that of the applied magnetic field. The light scattered by ions is collected by the lens with a front focal length of 70~mm and passes a quarter waveplate (QWP) and polarization beam-splitter (PBS) set for the transmission one of the $\sigma$ polarizations. The coupling into a single-mode optical fiber defines the spatial observation mode. The collimated output beam is sent to a detector arrangement consisting of a 50:50 beam splitter and a single-photon avalanche diode (SPAD) in each of its output ports. In the same detection channel, EMCCD camera is optionally used for exact determination of the trapped ion number and spatial dimensions of the crystal before and after each measurement of the coherence of the scattered light. The total detectable count rate of photons scattered from a single trapped ion positioned at the focus of the observation optics corresponds to about 4500~counts/s. We have characterized the scaling of the detection efficiency as a function of the axial and radial ion positions to access the effective volume of the detection mode.
The estimated detection volume in the trap measured in-situ corresponds to approximately Gaussian profiles with a FWHM widths of about 180~$\mu$m and 3~$\mu$m in the transverse plane and along the detection optical axis, respectively. The resulting ion configurations correspond to an ion string with one to up to $763\pm 90$ ions in the trap, and to an ellipsoid with concentric shell crystal structure~\cite{totsuji2002competition}. Between these two regimes, we were able to optimize the trapping properties and ion number to reach the particular 3-d configurations where two and three ions are positioned nearly along the optical axis of the collection lens and could contribute with near equal probabilities to the detected optical mode. These settings were particularly convenient for the direct evaluation of the initial scaling of $g^{(2)}(\tau)$ with the number of ions $N$.

%\section{Second-order coherence of light interfering from many atoms}

\begin{figure}[th!]
\begin{center}
\includegraphics[width=1\columnwidth]{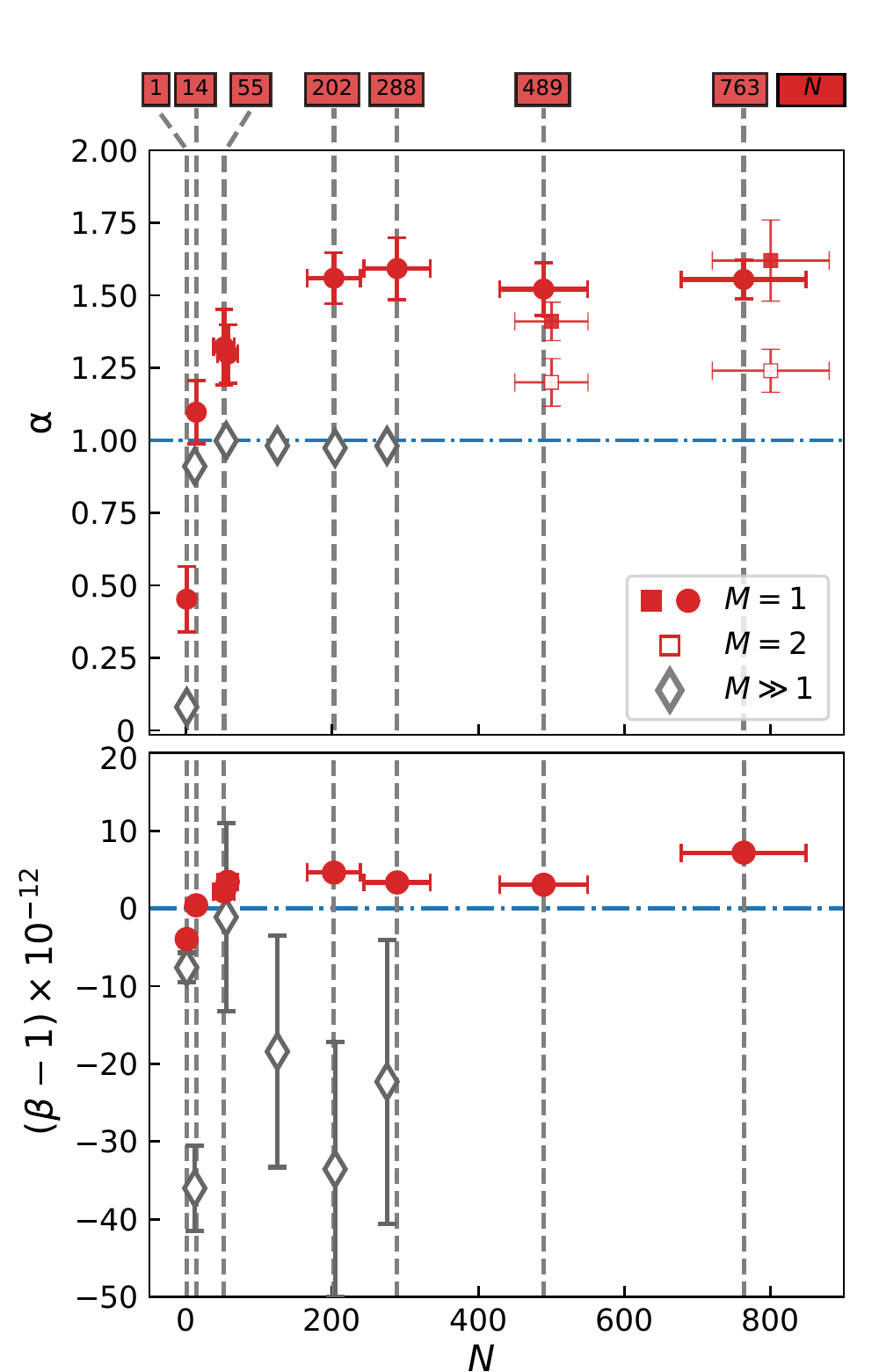}
\caption{The evaluated dependence of $\alpha$ and $\beta$ parameters for different numbers of ions $N$ in the trap. The measurements of the $\alpha$ from the single mode HBT detector are shown as red circles and were systematically acquired using the same observation conditions. The illustration of the effect of increasing the number of detection modes on the detected photon statistics is provided in measured data points shown as filled and empty red squares, which correspond to the detection of a one and two polarization modes, respectively. The spatially multimode data shown as grey diamonds were included from previous measurements in the same experimental setup~\cite{obvsil2018nonclassical}. The dash-dotted lines mark the provable threshold for unambiguous detection of super-Poissonian statistics of detected light fields corresponding to $\alpha>1$ and $\beta>1$, respectively. The uncertainties are evaluated statistically to a single standard deviation and are below the depicted symbol size where not shown.} % we should also include the N, i.e., number according to fit to the Eq. 7!}
\label{results}
\end{center}
\end{figure}

%$d\gg \lambda$

We approach the second-order coherence of light scattered from atoms in a single-mode detection regime. In order to obtain a phenomenological insight into the experimentally realized process, let's consider the scattering of a coherent beam from temporarily stable ensemble of $N$ identical atoms and with a large thermal motion corresponding to a position uncertainty $\langle \mathbf{r}_i^2 \rangle \approx \lambda$ achieved by the Doppler cooling settings far from the optimal regime, where $\lambda$ is the wavelength of the scattered light. Within these limits, the second-order correlation function $g^{(2)}(\tau)$ can be used to investigate an exclusive scenario where a response of two photon detectors on many emitters directly reflects the mutual indistinguishability and coherence of light scattered from different emitters. The time-delay $\tau$-dependent second-order correlation function can be defined in the normal ordering as
\begin{equation}
G^{(2)}(\tau)=\langle a^{\dagger}(0)a^{\dagger}(\tau)a(\tau)a(0) \rangle
\label{g2def}
\end{equation}
with a time dependence of the operators given by
\begin{equation}
a(\tau)\propto \sum_i \sigma_{i,-} (\tau) e^{-i \boldsymbol{k} \cdot \boldsymbol{r}_i}.
\end{equation}
where $\sigma_{i,- (+)}$ is the atomic lowering (raising) operator and $\boldsymbol{k}$ is the wave vector of the emitted light~\cite{loudon2000quantum}. We consider experiment in the limit of large fluctuations of the relative phase of light scattered from different emitters corresponding to large thermal position fluctuations of ions. We also focus solely on the investigation of the zero-time delay limit $\tau=0$, however, the derivation with a full temporal dependence can be found in the Appendix. Assuming statistically independent emission from individual identical emitters to the detected signal, the unnormalized second order correlation can be reexpressed as
\begin{equation}
\begin{aligned}
G^{(2)}(0)=\sum_i G_i^{(2)}(0)+\sum_{i \neq j}\lbrace G_i^{(1)}(0) \left[G_j^{(1)}(0)\right]^* +\bar{n}_i \bar{n}_j\rbrace,
\end{aligned}
\end{equation}
which consists of first-order $G_i^{(1)}(0)$ and second-order $G_i^{(2)}(0)$ correlation functions of individual emitters and their steady mean number of photons $\bar{n}_i$. When considering an optimal equal contribution of $N$ emitters to the detected signal, the normalized second order correlation function becomes
\begin{equation}
g^{(2)}(0)=\frac{\bar{g}^{(2)}(0)}{N}+\frac{N-1}{N}\left[\vert\bar{g}^{(1)}(0)\vert^2+1\right],
\label{eq:g2}
\end{equation}
where $\bar{g}^{(n)}(0)$ is $n$-th order normalized correlation function of a single emitter. The corresponding limiting cases for $N=1;2;3$ independent emitters give $g^{(2)}(0)=0;1;\frac{4}{3}$. The large but realistic finite emitter number limit, i.e. $N=100$, results approximately in $g^{(2)}(0)\rightarrow 2$. Crucially, reaching the super-Poissonian $g^{(2)}(0)>1$ in this scenario clearly requires a non-zero first order coherence $\vert\bar{g}^{(1)}(0)\vert > 0$. On the contrary, the fully incoherent emission corresponding to $g^{(1)}(0)=0$ approaches asymptotically $g^{(2)}(0)\rightarrow 1$ for a large $N$. We remind that the considered fundamental limit of independent atomic emitters effectively precludes the processes of scattered photon re-absorption or re-emission, which would effectively lead to temporal averaging of the correlation functions.

The analysis can be extended to cover measurement of $M$ independent photonic modes. It is relevant in situations where a detection process could potentially identify an emitting atom from an atomic ensemble or provide partial information about this atom. The photonic operators of individual modes then take form
\begin{equation}
a_{\kappa} \propto \sum_{i}u_{\kappa,i}\sigma_{i,+} e^{i \boldsymbol{k}\cdot \boldsymbol{r}_i},
\label{SM:RT}
\end{equation}
where $u_{\kappa,i}$ are amplitudes of a photon radiating from an atom with index $i$ into mode $\kappa$. The photon detector observes only some of the emitted spatial modes which we effectively consider by including the amplitudes $u_{\kappa,i}$ such that they preserve the normalization condition $\sum_{\kappa=1}^M |u_{\kappa,i}u^*_{\kappa,i}|=\eta$, where $\eta$ is the collection efficiency. We assume here for simplicity that the collection efficiency is the same for all atoms.
The multi-mode nature can be accounted for %the second-order correlation reads
%\begin{equation}
%g^{(2)}(\tau)=\frac{\bar{g}^{(2)}(\tau)}{N}+\frac{N-1}{N}\left[C\vert\bar{g}^{(1)}(\tau)\vert^2+1\right],
%\label{eq:g2C}
%\end{equation}
by including a numerical factor $C \in \left[0,1\right]$, which quantifies a degree of indistinguishably of emitted photons and provides the suppression of the impact of the first-order coherence $\bar{g}^{(1)}$ on $g^{(2)}$ due to partial distinguishability of the detected photons
\begin{equation}
\begin{aligned}
\vert\bar{g}^{(1)}(0)\vert^2 \rightarrow C \vert\bar{g}^{(1)}(0)\vert^2,\\
C=\frac{1}{\eta^2 N(N-1)}\sum_{i \neq j,\kappa,\lambda} u_{\kappa,i} u^*_{\lambda,j} u_{\lambda,i} u^*_{\kappa,j}.
\end{aligned}
\end{equation}
This allows us to capture realistic situations when a detector can partially recognize an emitting atom in an atomic ensemble. In the single-mode limit $C=1$ and first-order correlation $\bar{g}^{(1)}(0)$ affects significantly the second-order correlation $g^{(2)}(0)$. In the particular exquisite limit of emission of different ions in different polarization modes $C=0.5$. When a number of modes that the emitted light populates increases, $C$ decreases and can approach zero gradually. Note, that this is expected in cases of distinguishable emission, where light scattered from different ions at least partially populates different spatial modes. On the contrary, coherent emission in the identical mode superposition results in $C=1$, in accordance with indistinguishability of contributing ions and with the fundamental possibility of rotation of the modal basis to a single mode.
The full derivation of formula~(\ref{eq:g2}) including the multi-mode analysis can be found in the Appendix.

At considered zero time delay, the first term in the Eq.~(\ref{eq:g2}) vanishes in the case of an ideal single-photon emitters with $\bar{g}^{(2)}(0)=0$ and the equation qualitatively navigates to a broader class of interference effects that influence collective second-order coherence $g^{(2)}(0)$ purely by the first-order coherence $\bar{g}^{(1)}(0)$ of individual emitters. We note that in the limit of $N\rightarrow \infty$, the resulting super-Poissonian light shares this characteristics with chaotic light modelled by scattering of the coherent field from classical point scatterers with a random phase~\cite{loudon2000quantum}. However, the presented scatterers - atomic ions significantly differ from classical point reflectors. They constitute an intrinsic quantum emitters of nonclassical light - single photons~\cite{higginbottom2016pure}. Crucially, the ion trapping apparatus allows for scaling of their number within a large but strictly finite limit. Contrary to the classical example, the statistics of light emitted from a finite ensemble of atoms must always be nonclassical~\cite{loudon2000quantum,klyshko1996observable,lachman2016nonclassical,moreva2017direct,obvsil2018nonclassical,qi2018multiphoton}.

%\section{Measurements with large ion crystals}

We present the observations of the second-order coherence and complementary statistical evaluation of light scattered from a large but finite number of non-interacting and statistically independent atoms in the regime where the mutual phases of scattered photons are incoherently modulated by thermal atomic motion. The evaluated statistical correlation parameters shown in the Fig.~\ref{results} summarize the measurements of their dependence on the number of scatterers in trapped 3-d ion crystals. Fig.~\ref{results}-a) depicts the dependence of the parameter
\begin{equation}
\alpha=P_{\rm c}/P_{\rm s}^2,
\end{equation}
which can be evaluated from directly measurable probabilities of steady single photon counts $P_{\rm s}$ and coincidences $P_{\rm c}$ found within a set coincidence window $T$ with time delay $\tau=0$ between the two detection events in the setup employing two SPADs~\cite{grangier1986experimental}. Directly, if $\alpha>1$, probability of detecting coincidence is higher than for Poissonian noise with same photon flux. In the high attenuation limit corresponding to a small overall detection efficiency of photon emitted by a single ion in this setup $\eta \approx {10^{-4}}$, the value of directly measurable $\alpha$-parameter approximates well the idealized measurement of the normalized second-order coherence $g^{(2)}(0)$, which is otherwise not exactly measurable with finite arrays of bare single-photon detectors~\cite{sperling2012true}. Further, $g^{(2)}(0)>1$ determines super-Poissonian statistics of emitted light, whereas $g^{(2)}(0)>g^{(2)}(\tau)$ proves the bunched light~\cite{zou1990photon}.
The values of red full circles in the Fig.~\ref{results}-a) are measured in the single-mode detection regime $M=1$. They increase monotonically with the total number of ions in the crystal $N_{\rm tot}$, in qualitative agreement with predictions of the Eq.~(\ref{eq:g2}). The observed strongly sub-Poissonian $\alpha=0.4\pm 0.1$ limited by the finite detection time uncertainty of the employed SPADs specified to about 1~ns unambiguously certifies the nonclassical close-to a single-photon emission regime of a single ion. We note, that $\alpha$ value could be readily improved in our setup to $\alpha=0.071\pm 0.032$ by reducing the excitation power of the 397~nm laser, or even to $\alpha=0.036\pm 0.012$ in the pulsed excitation scheme, which doesn't suffer from the timing jitter of employed SPADs~\cite{higginbottom2016pure,obvsil2018nonclassical}. %Here $s_{397}=...$ is the off-resonant saturation parameter and $\delta$ corresponds to the detuning of the 397~nm excitation laser from the resonance.
The time bin in the presented evaluations has been optimized with respect to observable temporal features of $\alpha$ and noise to $T=1$~ns.
The possibility of pulsed measurements~\cite{higginbottom2016pure,crocker2019high,obvsil2018nonclassical} in the single-mode detection arrangement has been ruled out by relatively small detection efficiency of fluorescence collected in a single spatial mode per excitation pulse, which together with the achievable minimal time of the excitation sequence lowers the observable photon rate by more than an order of magnitude. We emphasize that the data for a single ion case were acquired for the 397~nm excitation laser frequency tuned closer to the resonance in order to enhance the photon rates and thus the feasibility of photon correlations, while the Doppler cooling conditions were not very crucial in this measurement. However, all data for number of ions in the trap $N>2$ were taken at same excitation conditions corresponding to significantly lower saturation parameter in order to further suppress the negative effect of time-jitter on the observations of photon statistics in a finite small time window and to maximize the temporal stability of the crystal in the limit of large ion crystals. % and corresponding large detunings of \textbf{$\delta_{397}\sim XX$~MHz in order to maintain the large atomic motion, which is crucial for the effective phase averaging}.
While the minimization of the statistical uncertainty of measured $\alpha$ still corresponded to a crucial optimization task, larger ion numbers naturally improve the photon rates in the same detection light mode. The transition from the sub-Poissonian to the super-Poissonian regime with $\alpha>1$ is clearly observable for the crystal with $N = 14$, where two ions were aligned to contribute equally to the observation spatial mode. For higher ion numbers $N \gg 202$, the $\alpha$-parameter further reaches about $\approx 1.56$, where it saturates mostly due to a limited contribution of ions from the outer shells of the 3-d oblate spheroidal crystals.

The effect of coherent contribution from many ions to the same optical mode can be further highlighted by comparison with the multi-mode detection scenario. Fig.~\ref{results} includes a set of experimental data taken in the same ion trapping setup and with similar excitation conditions from our previous experiments focused on unambiguous observation of multi-mode photon correlation measurements and nonclassical statistical properties on light scattered from large ensembles of ions~\cite{obvsil2018nonclassical}. Crucially, the nonclassical statistical character is expected to be preserved in the single-mode case, however, its direct observability in the value of $\alpha$ is counter-intuitively overridden by the increase of the second-order coherence at small time delays. The paramount sensitivity to detection modeness can be experimentally manifested by inclusion of an additional polarization mode by removing the polarization filter in the photon detection setup. The measurements with $N \approx 500$ and $N \approx 800$ ions in the crystal shown as full and empty red squares, respectively, demonstrate a decrease of the observable super-Poissonian value by a factor of $(\alpha_{M=1}-1)/(\alpha_{M=2}-1)=2.3\pm 0.3$ in a close agreement with the theoretically evaluated expected ratio of~2. This strong dependence on the detected optical mode number is a direct manifestation of a challenging aspect of observability of second order coherence from many independent atoms.

While the $\alpha$-parameter can be conveniently employed for the estimation of the second-order coherence without any direct sensitivity to photon losses, the parametrization of the measured photo-click statistics by the normalized parameter
\begin{equation}
\beta=P_{00}/P_{0}^2
\end{equation}
can be used for the dual evaluation of light statistics in the form which is insensitive to an independent Poissonian noise and, contrary to the $\alpha$, sensitive to optical attenuation~\cite{lachman2016nonclassical,lachman2022quantum}. Here $P_0$ and $P_{00}$ are the probabilities of no photon detection event for the given time bin on a particular SPAD and the probability of no photon detection on both SPADS with delay time between evaluated time bins set to $\tau=0$, respectively. Directly, if $\beta>1$, probability of a no-detection coincidence is higher than for Poissonian noise with same photon flux. It becomes clear that, although these probabilities can be calculated from probabilities of photon detection $P_{\rm s}, P_{\rm c}$, the resulting dependencies in the two different representations are very useful for a recognition of some of the complementary features of the detected light~\cite{lachman2016nonclassical,obvsil2018nonclassical,qi2018multiphoton,moreva2017direct}. This is important for our experiment to understand the sensitivity of observed results to the effective uncorrelated Poissonian noise from some of the emitters. Crucially, the condition $\beta<1$ on nonclassicality of values of $\beta$-parameter can be also derived ab-initio, as for $\alpha$-parameter, for the realistic detection system without any prior assumptions about the rate or efficiency of the photon detection apparatus. Thus, they can be conveniently employed for characterization of light emission  in the unexplored limits of increasing photon rates from a large number of photon emitters. The full red points in the Fig.~\ref{results}-b) show the measured dependence of the $\beta$ on the total number of ions in the crystal $N$ in the single-mode and multi-mode regimes. The observed transition for the single mode case from the negative value for a single ion to positive values at $N=14$ and monotonously increasing up to $N=202 \pm 36$ ions further confirms that adding ions on average results in photon statistics very different from the Poissonian noise~\cite{lachman2016nonclassical}. Crucially, the observed simultaneous gradual increase of $\alpha$ in both detection scenarios present very different behaviour in $\beta$. In the limit of a small mean photon number, $\alpha$ for single-mode thermal light approaches $2-\overline{n}$, whereas $\beta\approx 1+\overline{n}^2/4$ behaves quadratically. The single mode measurements result in the gradual increase of coincidences in the non-detection events on the two detectors, contrary to the decrease in the multi-mode detection case. This difference further underlines the fundamental tendency of the scattered light and, complementarily, photon non-detection events, to generate super-Poissonian light due to the effective indistinguishability of the contribution from different ions in the single-mode case. At the same time, the multi-mode scenario allows for the clear observation of the nonclassical character in $\beta$ with increasing dependence of the distance to the nonclassical threshold on the total number of ions in the trap $N$~\cite{obvsil2018nonclassical}. We note, that the presented single and multi-mode detection scenarios should be mostly considered for the understanding and emphasis of the conceptual behaviour of the coherence and statistical behaviour of the light emitted from ion crystals, rather than a direct quantitative comparison of the evaluated statistical parameters, as the actual $\alpha, \beta$ values are expected to be strongly dependent on the relative amplitudes of contributions of different ions in the trap. The error bars presented in the Fig.~\ref{results} were evaluated statistically from several consecutive measurements for each number of ions.
We emphasize that the realized light source formed by a large 3-d ion crystal has been considered without any assumptions on the realistically unavailable knowledge of the relative contributions of independent atoms or optical noise present in the source and detection apparatus. However, the observed dependence of $\alpha$-parameter allows to set a threshold on the smallest number of single-photon emitters $N_{\rm min}$ necessary for the observation of the given $\alpha$-values for the known total number of ions in the trap $N$, when considering the contribution of the rest of ions with a statistics of uncorrelated Poissonian noise. Such analysis confirms that although the ratio of the $N_{\rm min}/N$ decreases with the total number of ions in the crystal due to the effectively decreasing relative contribution of ions positioned far from the focus of the collection lens, $N_{\rm min}$ must gradually increase to up to $N_{\rm min}=189$ of independent single photon emitters necessary for explanation of the observed super-Poissonian values. The corresponding full analysis can be found in the Appendix.

%\section{Conclusions}

The observation of the super-Poissonian light scattered from a large, however fixed number of non-interacting and phase-randomized single-photon emitters, indistinguishable in a single optical mode, opens an exclusive experimental regime for studies of quantum optics of light-matter interaction. The corresponding practical demonstrations in the optical domain have been so far precluded mainly due to the mutually competing experimental requirements on the high probability of emission into the single mode from different emitters and on the non-interacting regime guaranteed by a large mutual distance. The latter limits the achievable detectable photon rates from individual scatterers and the corresponding possibility of observation of multi-photon coincidence detection events necessary for the second-order and higher-order coherence. We demonstrated feasibility of such implementation with large trapped ion crystals by an unambiguous observation of indistinguishable coherent contributions up to several hundreds of independent ions. %In comparison with the multi-mode observation case, the resulting statistics resulted in a clear super-Poissonian behaviour with gradually increasing value of the $\alpha$-parameter. Importantly, this transition emphasizes the new regime of controllable collective light scattering where mutually phase-incoherent but indistinguishable scatterers manifest in the coherent contribution to the observable photon correlations. We confirmed that the scattered field statistics follows the conceptual predictions for the emission from independent single photon emitters and demonstrated the crucial sensitivity to the number of observation optical modes.
The presented experimental observations are crucial for the realization of control of coherence of light upon the interaction with many independent atoms and will provide feasibility of emulations of complex quantum collective scattering phenomena~\cite{ficek2005quantum,thiel2007quantum,piovella2021classical,richter2022collective,bromley2016collective,Maiwoger2022,prasad2020correlating}. The bare observation of indistinguishability is fundamental for the implementation of photonic entangling schemes in the limit of a large number of ions~\cite{slodivcka2013atom}, experimental verification of super-Poissonian light from superradiance~\cite{bhatti2015superbunching,ficek2018highly}, exploitation of the nontrivial nonlinear coherent interactions of light and atoms~\cite{vogel1985squeezing,guerin2017light,inouye1999superradiant,corzo2019waveguide,prasad2020correlating}, increasing efficiency of nonlinear effects~\cite{manceau2019indefinite,spasibko2017multiphoton}, or schemes for enhancing efficiency of light-atom coupling and its coherent control~\cite{araneda2018interference,richter2022collective}.

\section*{Acknowledgements}

L. S., and D. B. are grateful for national funding from the MEYS under grant agreement No. 731473 and from the QUANTERA ERA-NET cofund in quantum technologies implemented within the European Union’s Horizon 2020 Programme (project PACE-IN, 8C20004). A. K., L. L., and R.F. acknowledge the support of the Czech Science Foundation under the project GA21-13265X. P. O., T. P., A. L., and O. C. are grateful to the project CZ.02.1.01/0.0/0.0/16\_026/0008460 of MEYS CR. L. P. acknowledges the internal project of Palacky University IGA-PrF-2022-005.

%\section*{Disclosures}
%
%The authors declare no conflicts of interest.

\clearpage

\newpage

%
%%%%%%%%%%% Prefix a "S" to all equations, figures, tables and reset the counter %%%%%%%%%%
\renewcommand{\theequation}{S\arabic{equation}}
\renewcommand{\thefigure}{S\arabic{figure}}
\renewcommand{\thesection}{S\arabic{section}}
\renewcommand{\theHequation}{Supplement.\theequation}
\renewcommand{\theHfigure}{Supplement.\thefigure}
\renewcommand{\bibnumfmt}[1]{[S#1]}
\renewcommand{\citenumfont}[1]{S#1}
\setcounter{equation}{0}
\setcounter{figure}{0}
\setcounter{table}{0}
\setcounter{section}{0}
\setcounter{page}{1} \makeatletter
%
%
%\section{Appendix}

\section*{Supplementary information}

\subsection{Emission from $N$ atoms to a single mode}

We derive the second-order coherence of light scattered from many atoms in a single-mode detection regime. We consider the scattering of a coherent light from a stable ensemble of $N_{\rm tot}$ identical ions. Their thermal motion corresponds to a position uncertainty $\langle \mathbf{r}_i^2 \rangle \approx \lambda$ or greater, where $\lambda$ is the wavelength of the scattered light. A single-mode time-delay $\tau$-dependent second-order correlation function is defined in the normal ordering as
\begin{equation}
G^{(2)}(\tau)=\langle a^{\dagger}(0)a^{\dagger}(\tau)a(\tau)a(0) \rangle.
\label{g2defS}
\end{equation}
A time dependence of the operators is given by
\begin{equation}
a(\tau)\propto\sum_i \sigma_{i,+} (\tau) e^{-i \boldsymbol{k} \cdot \boldsymbol{r}_i}.
\end{equation}
where $\sigma_{i,+ (-)}$ is the atomic raising (lowering) operator. The correlation function can be then rewritten as
\begin{eqnarray}
&& \langle a^{\dagger}(0)a^{\dagger}(\tau)a(\tau)a(0)\rangle \propto \nonumber \\
&&\langle \left(\sum_i \sigma_{\boldsymbol{r}_i,+}(0) \sigma_{\boldsymbol{r}_i,+}(\tau) +\sum_{i \neq j} \sigma_{\boldsymbol{r}_i,+}(0) \sigma_{\boldsymbol{r}_j,+}(\tau)\right)\times \nonumber \\
&&\left(\sum_i \sigma_{\boldsymbol{r}_i,-}(\tau) \sigma_{\boldsymbol{r}_i,-}(0) +\sum_{i \neq j} \sigma_{\boldsymbol{r}_i,-}(\tau) \sigma_{\boldsymbol{r}_j,-}(0)\right)\rangle,
\end{eqnarray}
where $\sigma_{\boldsymbol{r}_i,+}(\tau)=\sigma_{i,+}(\tau)e^{i \boldsymbol{k}\cdot \boldsymbol{r}_i}$ and $\sigma_{\boldsymbol{r}_i,-}(\tau)=\sigma_{i,-}(\tau)e^{-i \boldsymbol{k}\cdot \boldsymbol{r}_i}$ are used to abbreviate the notation. Expanding the brackets yields a summation of terms
\begin{equation}
\langle \sigma_{q,+}(0)\sigma_{r,+}(\tau)\sigma_{s,-}(\tau)\sigma_{t,-}(0)e^{i \boldsymbol{k}\cdot (\boldsymbol{r}_q+\boldsymbol{r}_r-\boldsymbol{r}_s-\boldsymbol{r}_t)}\rangle,
\end{equation}
which vanish when $\boldsymbol{r}_q+\boldsymbol{r}_r-\boldsymbol{r}_s-\boldsymbol{r}_t \neq 0$ due to a large fluctuation of the relative phase of light scattered from different emitters with large thermal position fluctuations. Considering only the non-vanishing terms results in
\begin{eqnarray}
&\ & \langle a^{\dagger}(\tau)a^{\dagger}(0)a(\tau)a(0)\rangle =  \nonumber \\
&\ &\sum_q \langle \sigma_{q,+}(0)\sigma_{q,+}(\tau)\sigma_{q,-}(\tau)\sigma_{q,-}(0) \rangle\nonumber \\
&+& \sum_{q \neq r} \langle \sigma_{q,+}(0)\sigma_{r,+}(\tau)\sigma_{q,-}(\tau)\sigma_{r,-}(0) \rangle\nonumber \\
&+& \sum_{q \neq r} \langle \sigma_{q,+}(0)\sigma_{r,+}(\tau)\sigma_{q,-}(\tau)\sigma_{r,-}(0) \rangle.
\end{eqnarray}
Assuming independent emission from individual identical emitters, it is then straightforward to express the unnormalized second-order correlation function as
\begin{equation}
G^{(2)}(\tau)=\sum_i G_i^{(2)}(\tau)+\sum_{i \neq j}\lbrace G_i^{(1)}(\tau) \left[G_j^{(1)}(\tau)\right]^* +\bar{n}_i \bar{n}_j\rbrace,
\end{equation}
which consists of first-order $G_i^{(1)}(\tau)$ and second-order $G_i^{(2)}(\tau)$ correlation functions of individual emitters and steady mean number $\bar{n}_i$ of photons for emitter with index $i$. When considering an ideal equal contribution of the emitters to the detected signal, the normalized second order correlation function becomes
\begin{equation}
g^{(2)}(\tau)=\frac{\bar{g}^{(2)}(\tau)}{N}+\frac{N-1}{N}\left[\vert\bar{g}^{(1)}(\tau)\vert^2+1\right],
\label{eq:g2S}
\end{equation}
where $\bar{g}^{(n)}(\tau)$ is $n$-th order correlation function of a single emitter. At zero time delay $\tau=0$, the equation~\ref{eq:g2S} further simplifies as the first term vanishes in the case of an ideal single photon emitter with $\bar{g}^{(2)}(0)=0$.

\subsection{Emission from $N$ atoms to $M$ modes}

The impossibility of a detector to determine which atom in an ensemble has radiated a photon represents a regime in which a photonic state is emitted into a single-mode. To describe the transition from this single-mode case to the multi-mode limit, we consider $N$ atoms that radiate into $M$ orthogonal modes. In this case, the second-order correlation function is defined by
\begin{equation}
    G^{(2)}(\tau)=\sum_{\kappa,\lambda=1}^M \langle a_{\kappa}^{\dagger}(0)a_{\lambda}^{\dagger}(\tau)a_{\lambda}(\tau) a_{\kappa}(0)\rangle,
    \label{SM:G2S}
\end{equation}
where $a_{\kappa}$ and $a_{\kappa}^{\dagger}$ correspond to annihilation and creation operator acting on the $\kappa$-th mode. Exploiting the quantum regression theorem, we can express these photonic operators by
\begin{equation}
    a_{\kappa}(\tau)=\sum_{i}u_{\kappa,i}\sigma_{i,+}(\tau)e^{i \boldsymbol{k}\cdot \boldsymbol{r}_i},
    \label{SM:RT}
\end{equation}
where $u_{\kappa,i}$ quantifies an amplitude that a photon is radiated from atom $i$ into mode with index $\kappa$. The amplitudes $u_{\kappa,i}$ obey the normalization $\sum_{\kappa=1}^M |u_{\kappa,i}u^*_{\kappa,i}|=\eta$, where $\eta$ represents the collection efficiency being independent of which atom radiates a photon. Thus, the second-order correlation function can be written in terms of atomic raising and lowering operators and the position of individual atoms. Since we allow for the random movement of atoms that exceeds the wavelength of the emitted light, all the terms in $G^{(2)}$ that depends on the position of atoms vanish due to this atomic jittering, and therefore the second-order correlation function works out to be
\begin{eqnarray}
&\ & G^{(2)}(\tau) =  \nonumber \\
&\ & \sum_{i,\kappa,\lambda} |u_{\kappa,i}|^2|u_{l,i}|^2\langle \sigma_{i,+}(0)\sigma_{i,+}(\tau)\sigma_{i,-}(\tau)\sigma_{i,-}(0) \rangle\nonumber \\
&+& \sum_{i \neq j,\kappa,\lambda} u_{\kappa,i} u^*_{\lambda,j} u_{\lambda,i} u^*_{\kappa,j}\langle \sigma_{i,+}(0)\sigma_{j,+}(\tau)\sigma_{i,-}(\tau)\sigma_{j,-}(0) \rangle\nonumber \\
&+& \sum_{i \neq j,\kappa,\lambda} |u_{\kappa,i}|^2|u_{\lambda,j}|^2\langle \sigma_{i,+}(0)\sigma_{j,+}(\tau)\times \nonumber\\
&\times& \sigma_{j,-}(\tau)\sigma_{i,-}(0) \rangle,
\label{SM:G2sigmas}
\end{eqnarray}
which can be further simplified by employing the normalization $\sum_{k}|u_{k,i}|^2=\eta$. Thus, the formula (\ref{SM:G2sigmas}) can be expressed as
\begin{eqnarray}
&\ &G^{(2)}(\tau)=\sum_i G_i^{(2)}(\tau)+\sum_{i \neq j}\bar{n}_i \bar{n}_j\nonumber\\
&+&\sum_{i \neq j,\kappa,\lambda} \frac{u_{\kappa,i} u^*_{\lambda,j} u_{\lambda,i} u^*_{\kappa,j}}{\eta^2}\lbrace G_i^{(1)}(\tau) \left[G_j^{(1)}(\tau)\right]^*\rbrace,
\label{SM:G2photonicsOp}
\end{eqnarray}
where $G_i^{(1)}(\tau)$ ($G_i^{(2)}(\tau)$) is the first-order (second-order) correlation function measured on the light that the $i$th atom would emit without contribution of other atoms and $\bar{n}_i$ is the mean number of photons that this photonic state would exhibit. A simplifying conjecture that $G_i^{(1)}(\tau)$, $G_i^{(2)}(\tau)$ and $\bar{n}_i$ are identical for all atoms enables the normalized correlation function $g^{(2)}(\tau)=G^{(2)}(\tau)/\left(\sum_{\kappa} \langle a_{\kappa}^{\dagger} a_{\kappa}\rangle\right)^2$ to reach a form
\begin{equation}
g^{(2)}(\tau)=\frac{\bar{g}^{(2)}(\tau)}{N}+\frac{N-1}{N}\left[C \vert\bar{g}^{(1)}(\tau)\vert^2+1\right],
\label{SM:eq:g2}
\end{equation}
where $\bar{g}^{(1)}(\tau)$ and $\bar{g}^{(2)}(\tau)$ are the normalized first-order correlation function and the normalized second-order correlation function that light from an individual atom exhibits, respectively, and the parameter $C=\frac{1}{\eta^2 N(N-1)}\sum_{i \neq j,\kappa,\lambda} u_{\kappa,i} u^*_{\lambda,j} u_{\lambda,i} u^*_{\kappa,j}$ quantifies a degree of indistinguishably of emitted photons. In the single-mode limit, $C=1$ and first order correlation $\bar{g}^{(1)}(\tau)$ affects significantly the second-order correlation $g^{(2)}(\tau)$. On the contrary, when a number of modes that the emitted light occupies grows up, $C$ decreases and can approach zero gradually. This allows us to cover realistic situations when a detector can recognize partially an emitting atom in an atomic ensemble.

\subsection{Lower limit on the contribution of independent atoms in the presence of uncorrelated noise}

\begin{figure}[t!]
\begin{center}
\includegraphics[width=1\columnwidth]{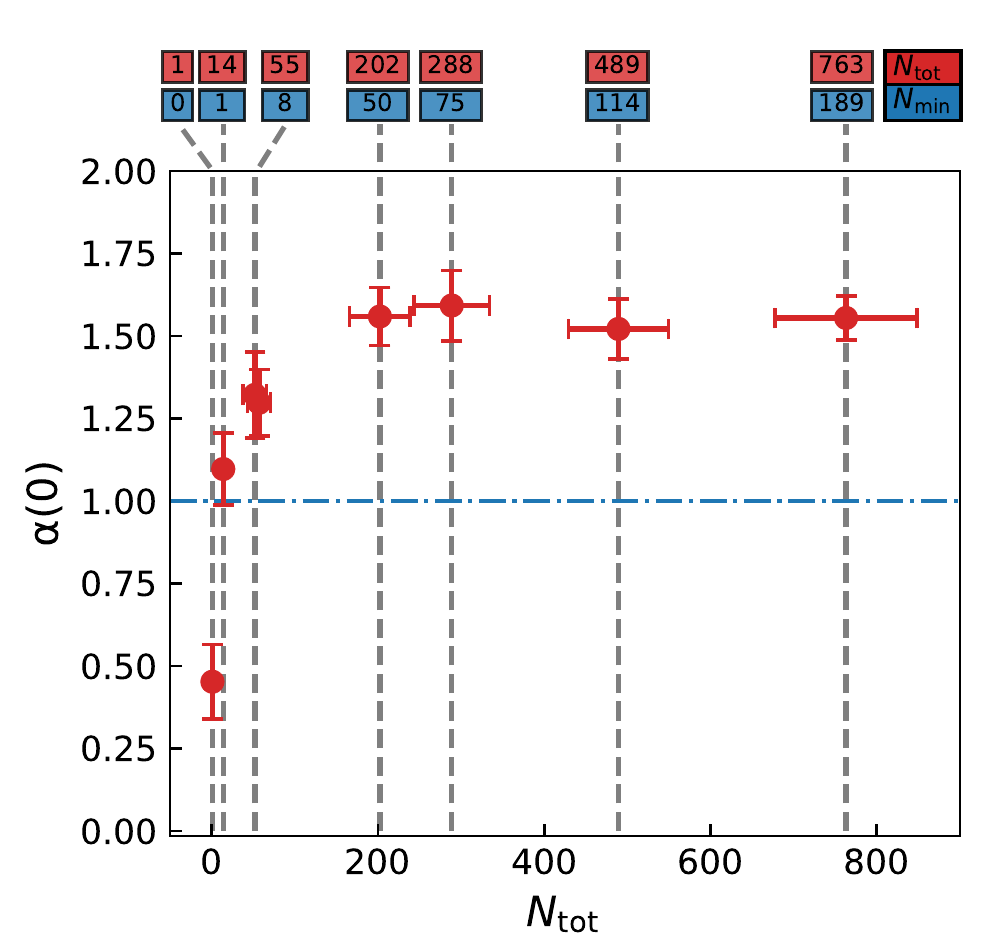}
\caption{The evaluated dependence of $\alpha(0)$ parameters for different numbers of ions $N_{\rm tot}$ in the trap including the evaluation of the threshold based criteria, which provide the limit on the smallest number of single-photon emitters $N_{\rm min}$, depicted on the top axis in blue squares.}
\label{results}
\end{center}
\end{figure}

\begin{table*}[th!]
\begin{center}
%\begin{tabular}{ m{5mm}  m{2.8cm} m{2.4cm} m{1.8cm} m{2.7cm} }
\renewcommand{\arraystretch}{1.5}
\begin{tabular}{p{0.2\textwidth}>{\centering}p{0.2\textwidth}>{\centering}p{0.2\textwidth}>{\centering\arraybackslash}p{0.2\textwidth}}
%\begin{tabular}{|c|c|c|c|}
\hline
Image of a crystal &  N &  $\alpha$ &  ($\beta - 1) \times 10^{-12}$ \\

\hline\hline

\includegraphics[width=2.5cm, trim=0 0 0 -50]{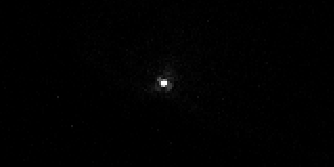} &  1 &  0.45 $\pm$ 0.11 &  $-3.94 \pm 0.82$ \\

\hline

\includegraphics[width=2.5cm, trim=0 0 0 -50]{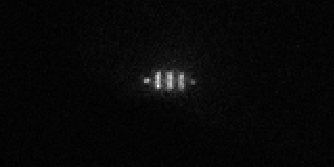} &  14 &  1.1 $\pm$ 0.1 &  0.41 $\pm$ 0.23  \\

\hline

\includegraphics[width=2.5cm, trim=0 0 0 -50]{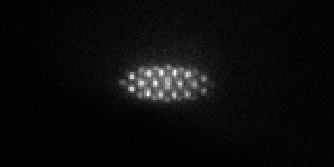} &  55 $\pm$ 14 &  1.30 $\pm$ 0.13 &  3.44 $\pm$ 1.00 \\

\hline

\includegraphics[width=2.5cm, trim=0 0 0 -50]{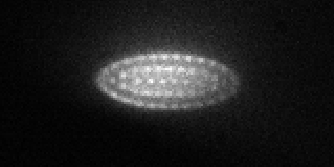} &  202 $\pm$ 36 &  1.56 $\pm$ 0.09 &  4.62 $\pm$ 0.69 \\

\hline

\includegraphics[width=2.5cm, trim=0 0 0 -50]{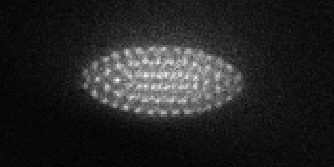} &  288 $\pm$ 45 &  1.59 $\pm$ 0.11 &  3.34 $\pm$ 0.59 \\

\hline

\includegraphics[width=2.5cm, trim=0 0 0 -50]{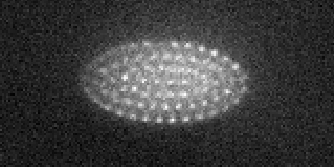} &  489 $\pm$ 60 &  1.52 $\pm$ 0.09 &  3.06 $\pm$ 0.51 \\

\hline

\includegraphics[width=2.5cm, trim=0 0 0 -50]{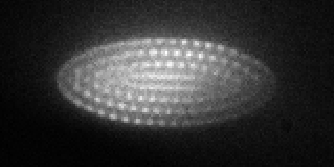} &  763 $\pm$ 86 &  1.56 $\pm$ 0.07 &  7.18 $\pm$ 0.78 \\
\hline
\end{tabular}
\caption{The list of measured spatial configurations of ion crystals including the corresponding estimated total number of ions and evaluated numerical values of $\alpha$ and $\beta$ parameters.}
\label{tbl:table_of_figures}
\end{center}
\end{table*}

The evaluated dependence of $\alpha$-parameter allows to set a threshold on the smallest number of single-photon emitters $N_{\rm min}$ necessary for the observation of the given $\alpha(0)$-value for the total number of ions in the trap $N_{\rm tot}$ when considering the contribution of the rest of ions $N_{\rm n}=N_{\rm tot}-N_{\rm min}$ with a statistics of uncorrelated Poissonian noise with a mean number of noise photons $\overline{n}= N_{\rm n} \eta$. Here, $\eta$ is an overall detection efficiency of photon emitted by an ion. Assuming merely independently emitting atoms, such analysis effectively excludes the possibility of observation of corresponding statistical dependence with small $N_{\rm min}$ on the order of a few ions. It allows for a simple conceptual approach to understanding the observed scattering behaviour with only basic description of the source and detection scheme, where the source constitutes of an ensemble of non-interacting atoms scattering the coherent light with near-equal probability for different atoms and the detection scheme considers spatially and polarization single-mode operation, which can be also unambiguously guaranteed in the experiment. The corresponding second-order coherence $g^{(2)}(0)$ then allows for a simple threshold based evaluation and comparison of the number of independent single-photon emitters $N_{\rm min}$ contributing indistinguishably to the detected signal with a single-photon signal with mutually uncorrelated phase.

Considering the probability of singles $P_{\rm s}=1/2 \eta N_{\rm tot}$ and probability of coincidences $P_{\rm c}=\eta^2 (\binom{N_{\rm min}}{2}+1/2 N_{\rm min} N_{\rm n}+(N_{\rm n}/2)^2$, the resulting
\begin{equation}
\alpha(0)=\frac{2 N_{\rm min}(N_{\rm min}-1)+2 N_{\rm min} N_{\rm n} +N_{\rm n}^2}{N_{\rm tot}^2}.
\end{equation}
It further reduces to
\begin{equation}
\alpha(0) \approx 1+(N_{\rm min}/N_{\rm tot})^2
\label{eq:g2withNoiseS}
\end{equation}
in the limit of $N_{\rm min}\gg 1$ applicable to the presented experimental regime for data points measured with $N_{\rm tot}>55$.

%The last summation in (\ref{SM:G2photonicsOp}) depends on the amplitudes $u_i$. This term contribute significantly in the single-mode limit. On the contrary, it vanishes gradually when number of mode that the emitted light occupies grows up. This allows us to capture realistic situations when a detector can recognize partially an emitting atom. In the limit of emission into many modes, this term becomes zero, and therefore the first order correlation $G_i^{(1)}(\tau)$ of individual atoms does not affect the second order correlation $G^{(2)}$.

\subsection{Measured $\alpha$ and $\beta$-parameters and corresponding configurations of ion crystals}

The table~(\ref{tbl:table_of_figures}) summarizes numerical values of evaluated statistical parameters for the data points presented in the main part of the manuscript. The corresponding ion crystal configurations are shown as an image from EMCCD and the number of trapped ions varies form a single to up to the $N=763\pm 86$. Each single experiment with constant number of ions in the Coulomb crystal represents collection of time-tagged photon counts from up to 12~hours long measurements on the correspondent trapped ion crystal. The evaluated parameter $\alpha=P_{\rm c}/(P_{\rm s})^2$, is evaluated from probabilities of steady single photon counts $P_{\rm s}$ and coincidences $P_{\rm c}$ with a zero time delay between the two detection time windows of a length $T=1$\,ns. The values of $\alpha$ and $\beta$ parameters and the corresponding uncertainties are evaluated from a set of six measured data points.
For the measurement with a single ion, the observed value $\alpha(0)= 0.45 \pm 0.11$ is limited by the finite detection time uncertainty of the employed SPADs ($\sim 1$\,ns) and the width of the anti-bunching dip given mostly by a high saturation parameter on the $4^2{\rm S}_{1/2}\leftrightarrow 4^2{\rm P}_{1/2}$ corresponding to the excitation with the 397~nm laser beam set for the generation of high fluorescence emission rate.

%\begin{figure*}[h]
%\begin{center}
%\includegraphics[width=1.8\columnwidth]{Figures/Fig_appendix.pdf}
%\caption{Example of the measured $\alpha(\tau)$ for given ion numbers and configurations corresponding to Columb crystal shown in the respective insets. The data were processed with the temporal bin resolution of 1~ns.}
%\label{fig:block_scheme}
%\end{center}
%\end{figure*}

\end{document}